\documentclass[aps,prb,twocolumn,groupedaddress,showpacs]{revtex4}
\usepackage{mathrsfs}
\usepackage{graphicx}  
\usepackage{cases}

\begin{document}

\title{Monte Carlo simulation based on dynamic disorder model in organic semiconductors:
From bandlike to hopping transport}

\author{Yao Yao}
\affiliation{State Key Laboratory of Surface Physics and Department
of Physics, Fudan University, Shanghai 200433, China}
\author{Wei Si}\affiliation{State Key Laboratory of Surface Physics and Department
of Physics, Fudan University, Shanghai 200433, China}
\author{Xiaoyuan Hou}
\affiliation{State Key Laboratory of Surface Physics and Department
of Physics, Fudan University, Shanghai 200433, China}
\author{Chang-Qin Wu}
\affiliation{State Key Laboratory of Surface Physics and Department
of Physics, Fudan University, Shanghai 200433, China}

\date{\today}
\begin{abstract}
The dynamic disorder model for charge carrier transport in organic
semiconductors has been extensively studied in recent years.
Although it is successful on determining the value of bandlike
mobility in the organic crystalline materials, the incoherent
hopping, the typical transport characteristic in organic
semiconductors, cannot be described. In this work, the decoherence
process is taken into account via a phenomenological parameter, say
decoherence time, and the projective and Monte Carlo method is
applied for this model to determine the waiting time and thus the
diffusion coefficient. We find the type of transport changes from
bandlike to incoherent hopping with a sufficiently short decoherence
time, which indicates the essential role of decoherence time in
determining the type of transport in organics. We have also
discussed the spatial extent of carriers for different decoherence
time, and the transition from delocalization (carrier resides in
about $10$ molecules) to localization is observed. Based on the
experimental results of spatial extent, we estimate the decoherence
time in pentacene has the order of 1ps. Furthermore, the dependence
of diffusion coefficient on decoherence time is also investigated,
and corresponding experiments are discussed.
\end{abstract}


\maketitle

\section{introduction}

It has been long-termly known that, the two different types of
charge carrier transport, namely, incoherent hopping and bandlike
tunneling, coexist in organic semiconductors, and both of them play
an essential role simultaneously.\cite{intro,NM} At the very
beginning, when the conducting conjugated polymer was firstly found
in 1970s, the explanation for the underlying mechanism of the
electric conductivity was based on the picture of solitons and
polarons.\cite{SSH} It is the strong electron-phonon (e-p) coupling
in organic materials that makes the electrons or holes self-trap in
a lattice distortion. This charged polaron is the main carrier,
which is initially realized to be completely localized in an
individual molecule. In this context, it was smoothly concluded
that, the character of charge transport between organic molecules
should be incoherent hopping of polarons. However, in the last
decade, plenty of experiments reported that there is a region that
bandlike tunneling
works.\cite{negative1,negative2,negative3,negative4,NM} For example,
there exists an abnormal region around 100-200K that the mobility
decreases as temperature increases (i.e., negative temperature
coefficient of mobility). This indicates the type of carrier
transport in this region is bandlike and coherent,\cite{NM} as the
charge is found to be delocalized in a region of several molecules
(about 10 molecules from the experiment of electron spin
resonance\cite{length}) rather than a single molecule. This result
is somehow contradiction to the traditional understanding of
localized polarons. In addition, it was also stated by a recent
experiment on the ultrafast initial carrier dynamics that, the
intramolecular microscopic dynamics should be significant in the
charge transport.\cite{ultrafast} Up to now, to describe the
\textit{dualistic} coherent and incoherent transport is still an
open subject.

To understand the new experimental findings, a number of theoretical
works based on the microscopic models, such as dynamic disorder
model,\cite{troisi,troisi2,SF1,SF2,theory1,theory2,theory3,theory4,theory5}
Su-Schrieffer-Heeger (SSH) model,\cite{An} and Anderson
model,\cite{Anderson} were addressed to account for the coherent and
incoherent transport,\cite{SF2,theory1,theory2} and the coherent
length and diffusion constant\cite{Anderson} have also been studied.
Based on Holstein-Peierls model, Troisi and his coworkers highly
commended the mechanism of dynamic disorder.\cite{troisi2} They
proposed that, due to the scattering with phonons, the carrier will
be localized into a small region, and thus the mobility decreases.
Their theory may be applied to the temperature dependence of
mobility in organic crystal but has not comprehensively described
that in organic semiconductors.\cite{troisi3} Meanwhile, under
sufficiently long time evolution, the model seems failing to capture
the basic bandlike characteristic.\cite{SF1} More importantly, the
Ehrenfest method used by Troisi has some fundamental drawbacks, and
the typical one is that the superposition principle is
deviated.\cite{proj} This is because the decoherence process has not
been appropriately taken into consideration,\cite{proj} while for
the simulation of incoherent hopping transport in organic
semiconductors, the decoherence process should be of actual
importance.

In the common sense, the role of phonons is twofold: One is to
provide the energy to help the carrier hopping (hopping transport)
and the other is to scatter with carriers to obstruct them (bandlike
transport). These two mechanisms might coexist in organic
semiconductors, which makes the discussion on this issue very
complicated. In this work, we apply the decoherence process into
Troisi's dynamic disorder model\cite{troisi} and then embed it into
the Monte Carlo simulation to quantitatively evaluate the diffusive
coefficient within organic molecules. Our aim is to provide a
generalized and efficient way to merge the above two roles of
phonons into a single framework and find the transition between
bandlike and incoherent hopping transport. The paper is organized as
follows. The next section will be the introduction of the present
theory. The main results will be in the third section, and the final
section is for the conclusions.

\section{Interplay of incoherent hopping and coherent evolution}

In this section, we will introduce the present theory. As the basis,
incoherent hopping and thus the decoherence process during
intermolecular hopping will be firstly described by introducing a
phenomenological parameter, i.e., decoherence time. Then the details
of the calculating procedure of diffusion coefficient will be
described. At last, the dynamic disorder model will be introduced
into the theory.

\subsection{Incoherent hopping}

We first discuss the incoherent hopping, which is the motivation we
consider the process of decoherence in this work. Originally, in
order to study the incoherent hopping, B\"{a}ssler applied the
kinetic Monte Carlo simulation with the so-called Miller-Abrahams
(M-A) formula embedded.\cite{KMC} Based on the Gaussian disorder
model (GDM), he was able to evaluate the relationship between the
mobility of carriers and temperature, electric field, and energy
disorder. Especially, he found for the temperature dependence of
mobility a scaling $\sim \exp(-T^{-2})$, which is quite different
from the traditional understanding.\cite{KMC} Following him, there
are lots of work investigating the influence of carrier
density,\cite{Density} electron-electron
interactions,\cite{Interaction} and so on.\cite{Others} Meanwhile,
the other theories, such as the molecular dynamics,\cite{theory5}
the percolation theory,\cite{perco1,perco2} and those considering
the trapping mechanism,\cite{trapping1,trapping2} are applied
extensively to complement the discussions of this incoherent
hopping. Especially, the dynamic disorder model has already been put
into the Monte Carlo simulation under the displaced harmonic
oscillator approximation,\cite{theory5} which is somehow different
from the present treatment.

Fundamentally, the M-A formula, coinciding with the detailed
balance, is related to a single phonon process, such that the energy
difference between initial and final site must be comparable with
the highest energy of phonons that could effectively interact with
carriers.\cite{Emin} Hence, the hopping mechanism based on M-A
formula could be described like this:\cite{hopping} initially a few
of vibrational modes (phonons) are excited, such that the carriers
could be heated up to make the energy levels of two neighboring
sites coincide, and then the phonon transit its energy to the
carrier to assist it hopping. This mechanism correspondingly
provides an intuitive picture to understand the application of M-A
formula in the case of low carrier's density,\cite{scale} but in
common cases, the vibrational modes are quite diverse, which means
the single phonon process loses its efficiency. Hence, a converted
picture of incoherent hopping should be addressed, which is one of
the main subject in this work.

\subsection{Process of decoherence}

In organic materials, charge carriers move in a complicated
environment, which is mainly composed of the disorders of molecular
configurations and vibrations, and decoherence happens incessantly.
Normally, the high frequency modes of intermolecular vibration (fast
interaction) could be realized to be the source of decoherence,
while the lower ones (slow interaction) is referred to a thermal
reservoir as in dynamic disorder model discussed in
below.\cite{troisi} The typical decoherence time is of the order
$1$ps in organic materials with good conductivity,\cite{Anderson}
such as in C60, where the frequency of center-of-mass motion is
observed to be 1.2THz,\cite{C60} but in more amorphous materials,
the value should be smaller. The common method treating decoherence
in molecular systems is to add some fluctuation and dissipation to
the model and evaluate the corresponding Green's
function.\cite{time} Obviously, this method is invalid for the
dynamic disorder model, in which the motion of lattice sites must be
calculated within real spacetime. So here we will describe a more
efficient picture for the decoherence process,\cite{deco} which is
straightforward from a so-called coarse graining method.\cite{proj}

\begin{figure}
\includegraphics[angle=0,scale=0.4]{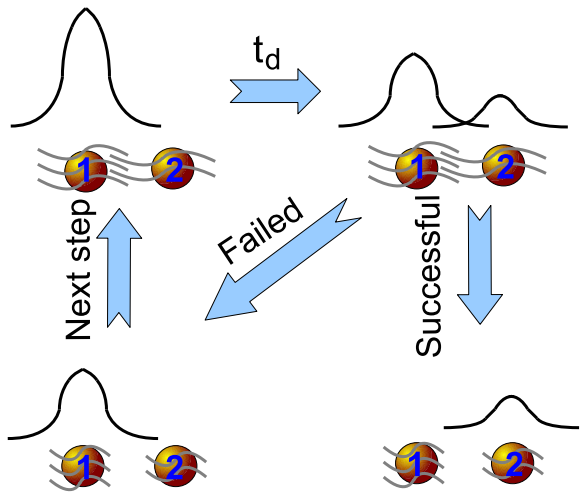}
\caption{Schematic for the process of incoherent hopping. Initially,
there is a polaron residing in molecule 1. After time $t_d$, its
wavepacket will spread to molecule 2. Then the correlation between
the two molecules is quenched due to the decoherence, and it will be
decided whether one try of the incoherent hopping is successful. If
yes, the polaron will hop to molecule 2, otherwise the whole process
restarts.}\label{decoherence}
\end{figure}

To this end, we take two molecules indexed by 1 and 2 as an example,
as shown in Fig. \ref{decoherence}. Initially, there is a carrier
residing in molecule 1 as a localized wavepacket. It will spread
following the equation of motion,
\begin{eqnarray}
\frac{\partial}{\partial t}\rho=-\frac{i}{\hbar}[H,\rho],\label{eom}
\end{eqnarray}
where we have used the tool of density matrix
$\rho\left(\equiv\left[\begin{array}{ccc}
\rho_{11}&\rho_{12}\\\rho_{21}&\rho_{22}\end{array}\right]\right)$
for convenience to understand decoherence. $H$ in (\ref{eom}) is the
Hamiltonian considering the simplest hopping term of polarons,
namely,
\begin{eqnarray}
H=\tau_{12}(C^{\dag}_1C_2+h.c.),\label{hamiE}
\end{eqnarray}
where $C^{\dag}_{i}$ ($C_{i}$) creates (annihilates) a polaron at
$i(=1,2)$-th molecule, $\tau_{12}$ the intermolecular overlap
integral of electronic wavefunctions.

When the frequency of vibration is high enough, the vibrational
modes make the decoherence happen very
quickly,\cite{picture1,picture2} such that we can define a
decoherence time $t_d$, within which the carrier will lose its
coherence and the off-diagonal terms $\rho_{12}$ and $\rho_{21}$
carrying the phase information of the system vanish.\cite{deco}
Normally, $t_d$ is of the order of $1/\tau_{1,2}$.\cite{picture2}
After $t_d$, the pure density matrix $\rho$ becomes a statistical
mixture, and due to the vanishing of overlap between 1 and 2, the
system could be realized to be a classical ensemble with $\rho_{11}$
and $\rho_{22}$ the probability that the carrier is residing in
molecule 1 and 2, respectively. One can then produce a random number
$r\in[0,1)$ as the procedure in Monte Carlo simulation. If
$r<\rho_{22}$, the carrier will be chosen to reside in molecule 2
after $t_d$, and one step of the incoherent hopping finishes.
Otherwise, the carrier remains at molecule 1, and the above process
restarts. This provides the generalized picture of decoherence.

\subsection{Evaluation of waiting time}

In the common studies of charge transport in organics, the coherent
motion of carriers could be simulated within both
adiabatic\cite{theory4} and nonadiabatic\cite{troisi,An} dynamical
method. As well, when one wants to study the incoherent hopping
only, the Marcus theory based on lattice relaxation and level
crossing is frequently used.\cite{marcus} Green's function was also
evaluated to distinguish incoherent and coherent
process.\cite{theory3} In this subsection, we will develop such a
method that merges both coherent and incoherent process on the basis
of decoherence as we described above. The central task is that, when
a carrier is initially residing in a molecule, we need to find a way
to determine the probability that a carrier hops out of the initial
molecule.

The procedure of our treatment is as follows. Initially, there is a
carrier localized at molecule $1$ (or the central site of the chain,
as discussed below), that is, the initial state of the system is
\begin{eqnarray}
|I\rangle=|1\rangle\equiv C^{\dag}_{1}|0\rangle.\label{ini}
\end{eqnarray}
The wavefunction of this carrier will then spread following
Eq.~(\ref{eom}). Since the carrier loses its coherence every $t_d$
as discussed above, it will choose a molecule (e.g., molecule $2$)
to reside in based on the probability at this molecule at $t_d$. It
means that, we calculate the probability of carrier at molecule $2$,
i.e., $P_2(t_d)(\equiv\langle t_d|C^{\dag}_2C_2|t_d\rangle)$ with
$|t_d\rangle$ the state of system at time $t_d$, and then make
$P_2(t_d)$ the hopping probability that the carrier hops out of the
initial molecule. This treatment is almost the same with the process
of decoherence as we discussed above.

Obviously, when $t_d$ is small enough, it is extremely difficult for
a carrier to hop out of the initial molecule by one try. So the key
step of the present theory is that, when the try fails, namely the
carrier keeps at the initial molecule, the wavefunction of system
will be projected onto the initial molecule, and continue evolving.
That is, a new round of evolution starts with the renewed projective
initial state
\begin{eqnarray}
|I_{new}\rangle=|1\rangle\langle 1|t_d\rangle.\label{initial}
\end{eqnarray}
Again we calculate $P_2(2t_d)$ after normalizing $|I_{new}\rangle$
by making the above steps recur. Of course, if necessary, $t_d$
could also be randomly selected. Now, we have in hand a series of
time-dependent $P_2(kt_d)$, which is different from those
time-independent hopping rates in GDM, so the remaining question is
to determine the waiting time $t_w$ for a carrier residing in the
initial molecule. Following the usual idea of Monte Carlo
simulation, we sum up $P_2(kt_d)$ and make the hopping happen when
the summation result exceeds a given random number, i.e.,
\begin{eqnarray}
\sum_{k=1}^{k_w}P_2(kt_d)>-\ln\xi,\label{inte}
\end{eqnarray}
where $\xi$ is a random number uniformly distributed between 0 and
1. In case for simplicity, the right hand side of (\ref{inte}) could
also be replaced by $1$. Then the waiting time is
\begin{eqnarray}
t_w=k_wt_d.
\end{eqnarray}
The above procedure could obviously be generalized for different
systems beyond two molecules. In the following, we will apply it to
the dynamic disorder model.

Before moving on to the next subsection, we would like to say that,
the projective method we describe here is almost the same with the
coarse graining method, with the only difference that the projection
is acting within the real space other than energy space.\cite{proj}
So that, the basic drawback of Ehrenfest method we will use shortly,
such as the deviation of superposition principle, is overcome in
some sense. Meanwhile, the sudden switching method of electronic
states in molecular dynamics is used for reference of the Monte
Carlo simulation.\cite{sudden} Equivalently, as we will show later,
one can also use a rate equation with the off-diagonal decay term to
calculate the diffusion process, which will give similar results
with the present theory in some case. But for a multi-site system,
since the velocities of decay of the off-diagonal term are not
exactly the same, the rate equation becomes quite complicated and
inefficient in practise. This is why we consider the projective
procedure here.

\subsection{Dynamic disorder model}

The dynamic disorder model recommended by Troisi is a quite simple
but efficient one in the study of charge transport in
organics.\cite{troisi,troisi2} The basic idea is to use a
one-dimensional SSH-like Hamiltonian to describe the coupling
between charge carriers and intermolecular vibrational mode. Since
the initial distribution of vibrational mode is dependent on
temperature, the disorder introduced by this electron-phonon
coupling is changeable with temperature, which implies the original
meaning of the words "dynamics disorder". The model Hamiltonian
writes,
\begin{eqnarray}
H=H_{\rm ele}+H_{\rm vib}.\label{hami}
\end{eqnarray}
The electronic part is
\begin{eqnarray}
H_{\rm ele}=-\sum_{j}[\tau-\alpha
(u_{j+1}-u_{j})](c^\dagger_{j+1}c_{j}+h.c.),
\end{eqnarray}
where $c^\dagger_{j}(c_{j})$ creates (annihilates) a carrier on the
$j$-th site. $u_j$ represents the displacement of the $j$-th site.
$\tau$ is the transfer integral and $\alpha$ is the electron-lattice
coupling constant. The vibrational part of Hamiltonian (\ref{hami})
is described as
\begin{eqnarray}
H_{\rm
vib}=\frac{K}{2}\sum_{j}(u_{j+1}-u_{j})^2+\frac{M}{2}\sum_{j}\dot{u}_j^2,
\end{eqnarray}
where $K$ is the elastic constant between neighbor sites and $M$ the
mass of a site. The parameters are chosen to the similar ones as in
[10]. That is, $K=14500$amu ps$^{-2}$, $M=250$amu,
$\alpha=100-995$cm$^{-1}/\rm{\AA}$, $\tau=30-300$cm$^{-1}$, and the
lattice constant is $4{\rm\AA}$. The number of site for the chain
will be set to sufficiently large, such that the diffusive
wavepacket of carrier can not reach the end of the chain.

Based on this Hamiltonian, we could use the Ehrenfest method to
calculate the diffusion of an initially localized wavepacket of
carrier.\cite{troisi} The initial state could have several forms and
result in slightly different trajectories, but the basic behavior
does not change.\cite{initial} In this work, the initial electronic
state we choose is to locate a carrier at a single site, e.g. the
central site of the chain. And the initial condition of vibrations
is under thermal equilibrium, i.e., $\{u_i\}$ and $\{\dot{u}_i\}$
will be randomly chosen in a Gaussian distribution with variance
$k_BT/K$ and $k_BT/M$, respectively, with $T$ the temperature. Then
we apply the time-dependent Schr\"{o}dinger equation for electronic
part and Newtonian equation for vibrational part to compute the time
evolution of the whole system.

In the original treatment of Troisi, the evolution time should be
long enough to ensure the saturation of diffusion coefficient. But
in the present work, we calculate the evolution for each round to
the time $t_d$ and then make a decision whether the carrier could
hop out of the initial site. The whole calculating procedure for
electronic part is almost the same as described in above
subsections. However, as we are now facing a multi-site system, we
treat the initial site as molecule 1 and the others as molecule 2,
and after the hopping is determined, we need to produce another
random number to determine which site the electron should go. On the
other hand, the evolution of vibrational part will be successive
during the decision of the final site of electron. The quantities we
compute are the diffusion coefficient defined as
\begin{eqnarray}
D=\langle r^2\rangle/2t_w,
\end{eqnarray}
with $r$ the distance between the initial site and the final site
that the carrier hops to, whose mean value is the diffusion length
as we will show in below. Of course, if necessary, the mobility
could be calculated by Einstein relation,\cite{troisi} but Einstein
relation is violated under nonequilibrium condition,\cite{Einstein}
such that the validity to use it here should be doubted. So we did
not plan to plot it in the present work.

\section{results and discussions}

\subsection{Comparison with no decoherence case}

\begin{figure}
\includegraphics[angle=0,scale=0.45]{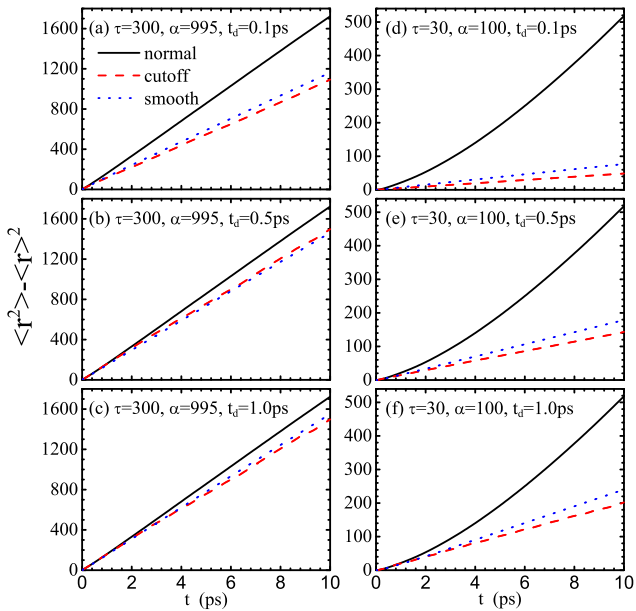}
\caption{Averaged squared displacement versus time for different
parameters with (red and blue) and without decoherence (black). The
red lines are from our projective method, while the blue ones are
from the calculation based on the master equation.}\label{troisi}
\end{figure}

In Fig. \ref{troisi}, we show the comparison between the present
theory and others. Firstly, under the parameters that Troisi has
used,\cite{troisi} we change the value of $t_d$ and calculate the
same quantity with Troisi, i.e., the averaged squared displacement.
The red lines are from our projective method, while the blue ones
are from the calculation based on the master equation. For the
latter case, we add an exponential decay term with variance $t_d$
for all the off-diagonal elements of the density matrix and
calculate the evolution and the physical quantities we want. This
treatment is an intuitive way to treat the decoherence at least in
two-site systems.

As shown in left panels of Fig. \ref{troisi}, following the $t_d$
decreasing, the diffusion becomes much slower than Troisi's results,
say no decoherence case. Meanwhile, when the $\tau$ and $\alpha$
reduce by one order, the reduction of diffusion coefficient becomes
quite large. These results clearly show that, when the decoherence
is presence, the ability of charge transport becomes poorer, and as
we will show shortly, the type transits from bandlike to hopping
transport. On the other hand, the results from the present
projective method are very close to those from the master equation,
but the computational efficiency improves a lot.

\subsection{Transition from bandlike to hopping transport}

\begin{figure}
\includegraphics[angle=0,scale=0.55]{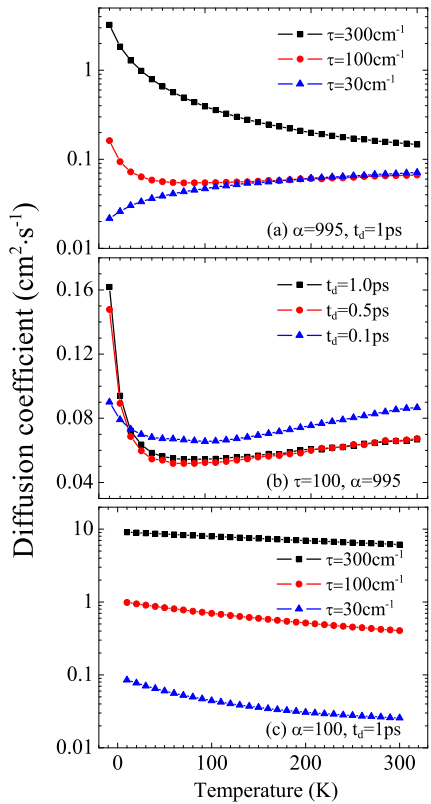}
\caption{Diffusion coefficient versus temperature for different
$\tau$, $t_d$, and $\alpha$. The values of $\alpha$ is in the unit
of cm$^{-1}/\rm{\AA}$. }\label{hopping}
\end{figure}
The most important characteristic difference between bandlike and
hopping transport is that, for the former one the diffusion
coefficient decreases with increasing temperature due to the thermal
induced dynamic disorder (say, a power law scaling $\sim T^{-m}$),
while for the latter case the diffusion coefficient increases with
increasing temperature due to the thermal assisted mechanism
($\sim\exp(-T^{-m})$). In Troisi's original studies, only bandlike
transport was found for all the range of parameters. Here in Fig.
\ref{hopping}, we show for different parameters the diffusion
coefficient. Our results clearly show that, when $\tau$ is small
($\tau=30$cm$^{-1}$, which is the typical value in organic small
molecules), there is only positive temperature dependence of
diffusion coefficient. While when $\tau$ is large
($\tau=300$cm$^{-1}$, which is the typical value in pentacene),
there is only a negative and power law ($m\simeq0.5$) temperature
dependence. In between ($\tau=100$cm$^{-1}$), both positive and
negative temperature dependence emerge. On the other hand, as shown
in Fig. \ref{hopping}(b), when we change $t_d$, we find a critical
region of the transition between bandlike (low temperature) to
hopping (high temperature) transport, and the transition point is
around 100K. This is the most important finding in this work. The
explanation of these results is that, for large $\tau$, the case is
the same with that discussed in dynamic disorder model,\cite{troisi}
while for small $\tau$, the carrier has little time to diffuse to
other sites, but some phonon induced potential well could assist
carrier to move. So in the latter case, temperature plays a positive
role, and the type of transport becomes incoherent hopping. More
interpretation will be addressed shortly when we study the diffusion
length. In addition, to get the sense of the influence of
electron-phonon coupling, we also show the results with
$\alpha=100$cm$^{-1}/{\rm \AA}$ in Fig. \ref{hopping}(c). Based upon
our calculation, the transition region from bandlike to incoherent
hopping is insensitive to $\alpha$ except some quantitative change,
which means the dynamic disorder itself does not influence the
transition of the two type of transport.

As we have mentioned, the temperature dependence is mostly negative
in crystalline organic semiconductors, whose mechanism should be
bandlike transport. Here, applying the physics of decoherence, we
obtain both positive and negative temperature dependence for
different $\tau$ and temperature. For the typical value of $\tau$ in
pentacene, namely $\tau=300$cm$^{-1}$, it shows negative temperature
dependence, which is the same with the result in dynamic disorder
model.\cite{troisi} It implies that, in this material, bandlike
mechanism dominates the charge transport, and the role of
decoherence is just to decrease the diffusion coefficient as we will
discuss later on. However, the value of $\tau$ in other organic
materials, such as Alq$_3$, is about two order smaller than that in
pentacene, i.e., $<30$cm$^{-1}$, so the temperature dependence
becomes positive. Hence, in common organic semiconductors under room
temperature, the characteristic of hopping transport should be more
frequently observed.

\subsection{Spatial extent of carriers}

\begin{figure}
\includegraphics[angle=0,scale=0.8]{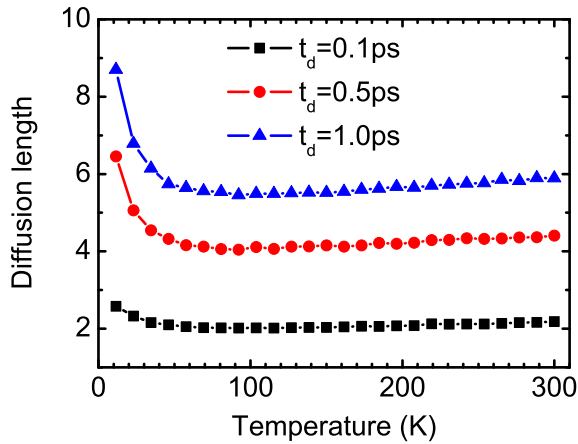}
\caption{Diffusion length (in the unit of the number of site) versus
temperature for different $t_d$. $\tau=100$cm$^{-1}$ and
$\alpha=995$cm$^{-1}/\rm{\AA}$}\label{dlength}
\end{figure}

The spatial extent of carriers is a controversial but crucial issue
in this field. Understanding of it is essentially helpful to clarify
the interplay between bandlike and hopping transport. The former
corresponds to a delocalized picture, while the latter is to the
localized one. Especially, since the localization length in
pentacene is estimated to be about $10$ molecules,\cite{length} this
value could be directly utilized to estimate the decoherence time in
pentacene. Within the present theory, we can also calculate the
diffusion length within $t_w$ as the average distance between
initial and final site that before and after carrier hops.
Obviously, when $t_d$ is very large, our result should be the same
with Troisi's.\cite{troisi} In Fig. \ref{dlength}, it is found that,
with $t_d\simeq0.5-1$ps, the carrier could diffuse over about $5$
sites, that is, the spatial extent of the carrier is about $10$
times intermolecular distance, which is very close to the
experimental prediction in pentacene.\cite{length} Hence, we can
safely say that, $1$ps is the typical order of decoherence time in
pentacene. Besides, with small $t_d$, the carrier is only localized
within the next-nearest sites, such that, hopping transport should
be the dominant for this case. Correspondingly, with temperature
increasing, the large $t_d$ curves decrease first and then increase
slightly, while the small $t_d$ curve does almost not change. This
means, for the bandlike transport, the spatial extent of carriers
should decrease with increasing temperature, while for the hopping
transport, the carrier is mainly localized. The results for other
parameters are similar, so we do not show here.

\begin{figure}
\includegraphics[angle=0,scale=0.4]{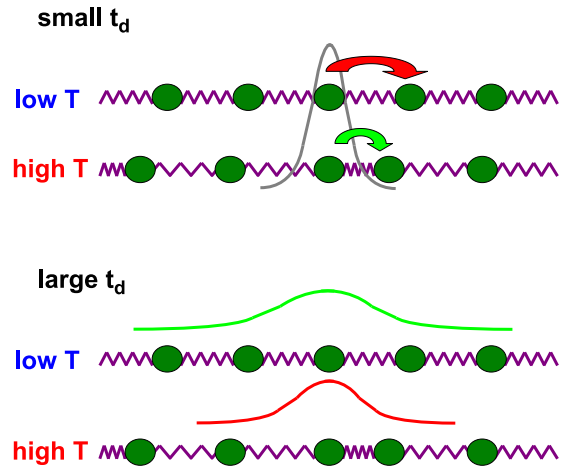}
\caption{Schematic for the transition between hopping and bandlike
transport. Under low temperature, the configuration of molecules is
regular, while under high temperature, it is disordered. For small
$t_d$, the carrier is mainly localized in one molecule, such that
temperature acts as an assistance. For large $t_d$, the wavepacket
of carrier is extended, and temperature produces disorders for the
carrier's diffusion. \label{trans}}
\end{figure}

The physical picture of these results is quite different from that
in the traditional understanding of incoherent hopping, where the
carrier residing in a trap would tunnel to another one with similar
energy.\cite{hopping} This mechanism works based upon the simple
phonon structure and dilute impurities in inorganic semiconductors.
In organic semiconductors, however, due to the dynamic disorder each
molecule might be treated as a trap and the phononic environment is
quite diverse. So the basic point for the dynamics disorder model
is, in our opinion, not the energy difference between molecules, but
how long the carrier spreads. Fig. \ref{trans} shows a brief
schematic for the present theory. When $t_d$ is small, the hopping
is mainly between two neighbor molecules. In this case, due to the
thermal motion of each molecule, the intermolecular distance
eventually decreases when the temperature is sufficiently high, such
that the hopping rate should be larger than that under low
temperature. This is actually the temperature assisted hopping.
While when $t_d$ is large, that is, the carrier is much easier to
spread out to several sites, the thermal motion of molecules behaves
now as the disordered obstacles. So in this case, temperature will
act to produce more and more scattering to the carriers and then
block the carrier's motion.

\subsection{Diffusion coefficient versus $t_d$}

\begin{figure}
\includegraphics[angle=0,scale=0.8]{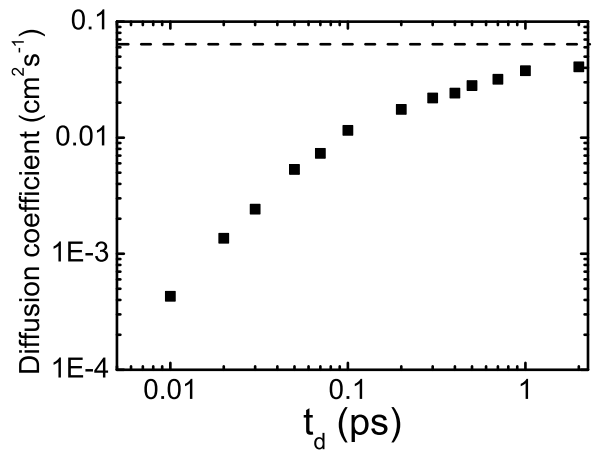}
\caption{Diffusion coefficient versus $t_d$. $\tau=300$cm$^{-1}$,
$\alpha=995$cm$^{-1}/\rm{\AA}$, and $T=300$K. The dashed line
denotes the diffusion coefficient obtained in Ref. [10].\label{td}}
\end{figure}

In the present theory, the decoherence time $t_d$ is the most
important parameter. In Fig. \ref{td}, we show the dependence of
diffusion coefficient on $t_d$. As expected, the diffusion
coefficient increases following the increasing $t_d$, and when
$t_d>2$ps, diffusion coefficient tends to saturate to the result
from Troisi.\cite{troisi} This is easy to understand, since when
$t_d$ is extremely small, it is quite hard for a carrier to hop out
of the initial site. It is worth noting that, the estimated value of
mobility in dynamics disorder model is slightly larger than that
from experiment, say $3$cm$^2$s$^{-1}$V$^{-1}$ from theory and
$1$cm$^2$s$^{-1}$V$^{-1}$ from experiment.\cite{troisi} The present
theory based on the decoherence might be the solution to this
difference. Furthermore, the experiment on organic field effect
transistor has shown that, when the drain voltage is large, the
temperature coefficient is negative, and for the inverse case, it is
positive.\cite{NM} This could also be explained in the present
theoretical framework. In case the drain voltage is large, the
waiting time $t_w$ in each molecule should be small. Considering the
ratio between $t_w$ and $t_d$ that, the smaller the $t_w$, the
larger the $t_d$, the relationship of mobility with drain voltage
could be easily understood.

\section{concluding remarks}

Actually, the influence of phonons on charge transport in organic
semiconductors has been studied extensively in the literature.
However, the meaning of "incoherent hopping" in organic materials is
still obscure in our opinion, since the widely used M-A formula is
not such applicable for organic molecules. Meanwhile, as we show in
this work, hopping and bandlike transport happens at different time
scales. The first case refers to decoherence time of the same order
with relaxation time of phonon so that phonon is able to transfer
its energy to the carrier to assist it hopping, while the second
case functions when $t_d$ is large enough to ensure the time for
scattering between carriers and phonons. In a real material, there
are many ways to adapt these two conditions and make the type of
transport transit from hopping to bandlike. Furthermore, under low
temperature, tunneling might become dominating, and carriers will
search for a molecule out of the trap which has closest energy to
tunnel to, i.e., Mott's variable range hopping
mechanism.\cite{hopping} This case has not been addressed in the
present work.

In the end, we would like to discuss more on the present theory.
Firstly, the practical device parameters, such as electric field,
have not been explicitly included, which could be easily considered
within the present framework. For example, the electric field could
be regarded as a phase factor in the hopping constant,\cite{phase}
but for the common Ehrenfest method, this way loses its efficiency
due to the Bloch oscillation. Especially, based on the study of
electric field, we might discuss the validity of Einstein relation
in these systems.\cite{Einstein} Secondly, it is straightforward to
replace the dynamic disorder model discussed in the present work,
such as Holstein model, spin-boson model, etc. Especially, since the
spin motion in organic materials is always realized to be coherent,
it could be easily incorporated into the present theory, and the hot
debated magnetic field effect could be studied.\cite{SpinBoson}
Thirdly, the theory could be applied to estimate the corresponding
decoherence time in the specific material, since it is the only
adjustable parameter.

\begin{acknowledgments}
We are grateful to S. Ciuchi, S. Fratini, T. Kreouzis, and W. Gillin
for valuable discussions. This work was supported by the NSF of
China, the National Basic Research Program of China (2009CB29204 and
2012CB921400), and the EC Project OFSPIN (NMP3-CT-2006-033370).
\end{acknowledgments}

\end{document}